# Jekyll RDF

## Template-Based Linked Data Publication with Minimized Effort and Maximum Scalability

Natanael Arndt[1,2][0000−0002−8130−8677], Sebastian Zänker[2],
Gezim Sejdiu[3][0000−0002−3441−1372], and Sebastian Tramp[4][0000−0003−4707−2864]

[1] AKSW, Leipzig University, Augustusplatz 10, 04109 Leipzig, Germany
arndt@informatik.uni-leipzig.de
http://aksw.org/NatanaelArndt
[2] Institut für Angewandte Informatik e.V., Goerdelerring 9, 04109 Leipzig, Germany
zaenker@infai.org
http://aksw.org/SebastianZaenker
[3] Smart Data Analytics, University of Bonn, Endenicher Allee 19a, 53115 Bonn, Germany
sejdiu@cs.uni-bonn.de
http://sda.tech/Person/GezimSejdiu/
[4] eccenca GmbH, Hainstr. 8, 04109 Leipzig, Germany
sebastian.tramp@eccenca.com
https://sebastian.tramp.name

**Abstract** Over the last decades the Web has evolved from a human–human communication network to a network of complex human–machine interactions. An increasing amount of data is available as Linked Data which allows machines to "understand" the data, but RDF is not meant to be understood by humans. With Jekyll RDF we present a method to close the gap between structured data and human accessible exploration interfaces by publishing RDF datasets as customizable static HTML sites. It consists of an RDF resource mapping system to serve the resources under their respective IRI, a template mapping based on schema classes, and a markup language to define templates to render customized resource pages. Using the template system, it is possible to create domain specific browsing interfaces for RDF data next to the Linked Data resources. This enables content management and knowledge management systems to serve datasets in a highly customizable, low effort, and scalable way to be consumed by machines as well as humans.

## 1 Introduction

In 2001 Tim Berners-Lee and James Hendler stated: *The Web was designed as an information space, with the goal not only that it should be useful for human–human communication, but also that machines would be able to participate and*

*help users communicate with each other* [6]. Now 18 years later we are at the point that a huge amount of data is published as Linked Data as it is apparent in the *Linked Open Data Cloud*[5] with 1,234 datasets and 16,136 links between the datasets[6]. But the RDF data is not suited and meant to be read and understood by humans. On the informal *Semantic Web Layer Model*[7] the top most layer represents *User Interface & Applications*. A great variety of applications exist to visualize RDF data. Such applications are table based triple explorers, like *pubby*[8], *LOD View*[9], and *LD Viewer/DBpedia Viewer* [17,18] and visual graph explorers like *LodLive*[10], *LODmilla* [20], and *Linked Data Maps* [21]. These applications are restricted to a view that is very close to the RDF data model and are thus suited for data experts who understand the concepts of RDF and the respective vocabularies, but not suitable for end users.

Web Content Management Systems (WCMS) are software systems to support the processes to create, manage, provide, control, and customize content for websites [11]. Besides the management of the content in a Content Repository, the customizable presentation using templating systems is a key aspect of WCMS. Semantic Content Management Systems (SCMS) extend Content Management Systems with additional functionality to enrich the content with semantic meta-data in a Knowledge Repository. Nowadays we are at the point that the semantic data is not "only" meta-data, but encodes the information itself. The activity to manage the semantic data as information by itself is called Semantic Data Management and gets a lot of attention [8,4]. To make this semantic data available to end users there is a need for semantic templating systems which is experiencing little research so far.

In this work, we present an approach for the generation of static Web exploration interfaces on Linked Data. The approach is based on devising a declarative DSL to create templates to render instance data. The templates are associated to RDF classes and a breath-first search algorithm determines the best-suitable template for any given data resource. To demonstrate the feasibility of the approach, we implemented it as an extension to the popular Jekyll[11] Static Site Generator and CMS[12]. In contrast to dynamic web pages, static web pages are preparatively generated and can be served without further server-side computation, thus providing highest possible scalability. This approach is complementary to the data focused approach of Linked Data Fragments[13] to reduce costly server-side request evaluation. By rendering RDF data to static HTML sites we provide a method to browse the Semantic Web seamlessly integrated with the rest of the Web and close the gap between structured data and human accessible exploration interfaces. To the best of our knowledge Jekyll RDF[14] is the first approach to apply the concept of Static Site Generators to RDF knowledge bases. In contrast to the state of the art (cf. section 2) it does not require programming knowledge of its user, does not need a dynamic back-end nor it is integrated in an IDE. It is provided as a standalone tool inspired by the UNIX (and more recently micro-service) philosophy to *make each program do one thing well* [12]. Because of the separation of concerns it is integrable with existing content management and knowledge management workflows. Due to the modular conception the presented method should be transferable to further Static Site Generators like Next, Hugo, and Hyde[15] or complex frameworks like Gatsby[16].

---

[5] http://lod-cloud.net/
[6] As of June 2018
[7] https://www.w3.org/2007/03/layerCake.svg
[8] http://wifo5-03.informatik.uni-mannheim.de/pubby/
[9] http://lodview.it, https://github.com/dvcama/LodView
[10] http://en.lodlive.it/, https://github.com/dvcama/LodLive
[11] https://jekyllrb.com/, https://www.staticgen.com/, https://www.netlify.com/blog/2016/05/02/top-ten-static-website-generators/
[12] https://www.siteleaf.com/, https://www.netlifycms.org/
[13] http://linkeddatafragments.org/

In this paper we first give an introduction to the state of the art in section 2. Then we provide an overview on the Static Site Generator architecture with detailed descriptions of the core components in section 3. An important aspect of the separation of concerns approach is the ability to integrate a tool with larger systems to accomplish high-level tasks. We present the integration of the Static Site Generator in a Linked Data tool chain in section 4. The Jekyll RDF system is already used in several setups from various domains which we present in section 5. Finally, we draw our conclusions and outline future work in section 6.

## 2  State of the Art

The generic data model provided by RDF allows the publication of data representing various domains and their aspects on the Web. The abstraction of the data model from its representation opens the possibility for arbitrary visualizations of the data. A great variety of systems exists that provide ways to access and visualize RDF data [16,13]. Many systems are created to serve a specific purpose such as visualizing a specific dataset or data expressed using a specific vocabulary. In the following we focus on frameworks and template based systems that provide a generic tooling to create custom exploration and visualization interfaces that are usable for any RDF dataset.

Templating systems usually provide a flexible approach for inserting data into a scaffolding of an HTML page. The *SPARQL Web Pages*[17] system defines a templating language that allows to incorporate data from an RDF graph into HTML and SVG documents. It is shipped with the commercial version of the *TopBraid Composer*. A similar approach is followed by *LESS* [5] which later was integrated with the *OntoWiki* [10]. The *OntoWiki Site Extension*[18] [9] allows to render RDF resources in HTML views using a PHP base templating language. To serve the required representation of a Linked Data resources the OntoWiki Linked Data server uses content negotiation to dynamically serve an HTML view to web browsers and an RDF representation to Linked Data systems.

---

[14] https://github.com/AKSW/jekyll-rdf
[15] https://nextjs.org/, https://gohugo.io/, http://hyde.github.io/
[16] https://gatsbyjs.org/
[17] http://uispin.org/
[18] https://github.com/AKSW/site.ontowiki

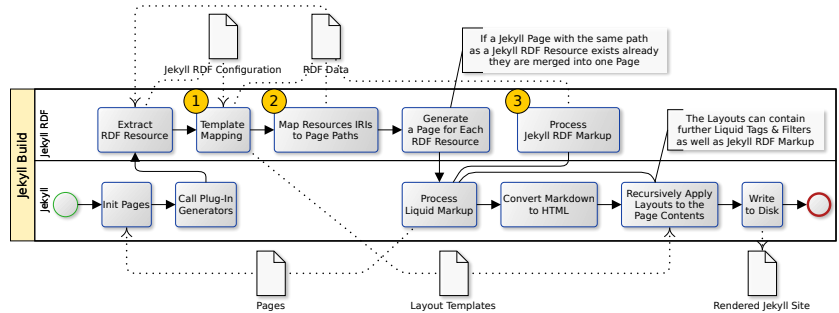

**Figure 1.** The architecture of Jekyll RDF and its interplay with Jekyll.

A different approach to provide customizable web interfaces to explore and even edit RDF data is presented by Khalili et al. with the *LD-R* [14]. It provides a framework to define *Linked Data-driven Web Components* in JavaScript. With this framework it is possible to reuse existing components and compose new dynamic web interfaces. A similar approach to build *Semantic Interfaces for Web Applications* is presented with the MIRA framework [7]. It defines an *abstract interface definition* that composes elements to form a hierarchy of widgets. These widgets can be used in JavaScript applications to build responsive user interfaces.

In summary, the related work of LD-R and MIRA [14,7] as well as complex frameworks like Gatsby aim at programmers and software engineers. The previous work of LESS and the OntoWiki Site Extension [5,10,9] provides a template based approach. But LESS and OntoWiki Site Extension as well as the application frameworks presented in [14,7] rely on a server side complex *dynamic* data management system. In this paper, our aim is to provide a *static* approach to maximize the scalability while minimizing the effort by following a templated based approach that can be used without programming knowledge.

## 3  The Static Site Generation Architecture

We conceived our system as an extension to existing Static Site Generators. A Static Site Generator usually translates a set of HTML and Markdown files (*pages*) to a collection of HTML files (*site*) by using structural *templates*. Pages and templates can be conditionally formatted and enriched with tags and filters defined by a markup language to embed data values and alter the presentation. The generated site is served either with an integrated web server or any other HTTP server. To showcase our approach we implemented Jekyll RDF[19] as a plugin for the popular Jekyll system.

The plugin system of Jekyll provides among others the possibility to add a *generator* to create new pages and to implement custom *tags* and *filters* for the Liquid markup language[20]. Jekyll RDF uses this system to provide its

---

[19] https://github.com/white-gecko/jekyll-rdf
[20] https://shopify.github.io/liquid/, https://shopify.github.io/liquid/basics/introduction/

**Listing 1.** The sections of the Jekyll configuration relevant for Jekyll RDF including base url, data source, selection of resources, and template mappings.

```yaml
1  baseurl: "/sachsen/"
2  url: "http://pfarrerbuch.aksw.org"
3  plugins: [jekyll-rdf]
4  jekyll_rdf:
5    path: "sachsen.ttl"
6    restriction: "SELECT ?resourceUri WHERE {?resourceUri ?p ?o . FILTER
           regex(str(?resourceUri), '^http://pfarrerbuch.aksw.org/sachsen/')}"
7    default_template: "resource"
8    class_template_mappings:
9      "http://xmlns.com/foaf/0.1/Person": "person"
10     "http://purl.org/voc/hp/Place": "place"
11     "http://purl.org/voc/hp/Position": "position"
12   instance_template_mappings:
13     "http://pfarrerbuch.aksw.org/": "home"
```

main functionalities as depicted in fig. 1: (1, 2) generate a Jekyll page for each resource from the RDF graph (cf. sections 3.1 and 3.2), and (3) extend the markup language by a set of filters and tags to query the RDF graph (*Jekyll RDF Markup Language*, cf. section 3.3). *Jekyll* controls the main flow of the build process which is depicted in the lower row. The particular tasks which are relevant for the rendering of RDF data are depicted in the upper row. The process needs several data artifacts namely, the *pages* to be rendered by Jekyll, the *configuration* options which are specific to Jekyll RDF, the *RDF data*, and the *templates* to defined the layout of the pages and RDF resources. The process to generate a Jekyll page for each RDF resource is split into four steps, extract the RDF resource from the RDF data model as specified in the configuration (cf. listing 1) and create program objects accordingly, map the resources to templates, map the IRIs of the RDF resources to according page paths, and generate a Jekyll page object for each RDF resource. The template mapping (no. 1 in fig. 1) can happen directly per RDF resource or based on the RDF types of a resource, this is described in detail in section 3.1. Design decisions required to represent the RDF resource's IRIs in the path system of Jekyll are explained in section 3.2 (no. 2). Further, Liquid is extended to the *Jekyll RDF Markup Language* which is presented in section 3.3 (no. 3).

In listing 1 an exemplary configuration file for Jekyll is provided with the relevant sections to configure a Jekyll RDF setup. Lines 1 and 2 together represent the URL under which a Jekyll site is served. In line 3 the Jekyll RDF plugin is registered. Lines 4 to 13 are the specific parameters to configure Jekyll RDF, the path (line 5) specifies the data source for the RDF data and the restriction (line 6) specifies the list of RDF resources to be rendered. Lines 7 to 13 specify the template mapping and are further described in section 3.1.

### 3.1 Template Mapping

Jekyll provides a templating system to allow the reuse of components on multiple pages and allow similar pages to look alike. The template for a Jekyll page is specified in a header in the page. During the rendering process Jekyll applies this template to the content of the page. In contrast to a Jekyll page an RDF resources has no "content" and no header and thus we need to specify the representation of the resource. In the following we introduce three mapping mechanisms to determine which template should be applied for a resource. The template assignment configuration is shown in lines 7 to 13 in listing 1. In the `instance_template_mappings` section each resource can be directly assigned to a template by specifying the resource IRI and the template name. Further, two indirect options to assign a template are specified. In the section `class_template_mappings` RDF classes are mapped to templates. Each resource that is an instance of a specified class or its subclasses gets this template assigned. The precedence order of the template mappings is: instance based, class based, `default_template`.

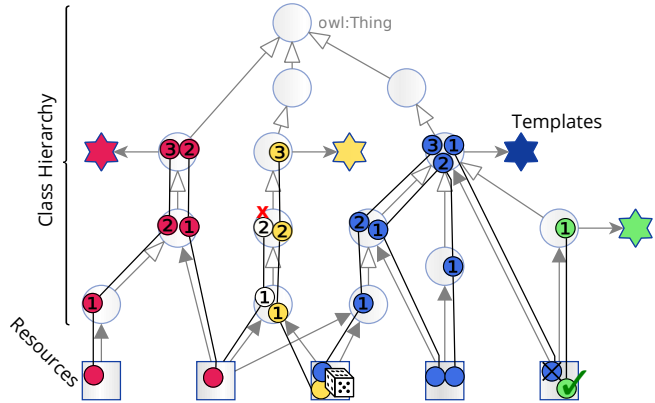

**Figure 2.** The class hierarchy is used to select the template to render a resource.

Other than for the instance based and default mapping the class template mapping introduces ambiguity as depicted in fig. 2. If a resource has multiple classes and each has a template assigned, it can not be decided which template to use for the resource. The template selection can not be limited to the trivial evaluation of `rdf:type` triples as this would not take the assignment of templates to super classes into account. Inferencing along `rdfs:subClassOf` relations would also be no good approach as it introduces more ambiguity and hides precedence information about the most specific class for a resource.

We decided to select the template for an instance according to three rules as depicted in fig. 2 (a *candidate* is a class that has a template assigned). (1) Select the closest candidate in the class hierarchy, (2) if more then one candidate exists with different templates but with the same shortest path distance, take the candidate with the most *specific* class, (3) if still no candidate could be selected, produce a warning and randomly select one of the candidates. A class $a$ is considered more *specific* than $b$ if an `rdfs:subClassOf` property path exists from $a$ to $b$ but not the other way around. To implement the template selection we chose a breath-first-search algorithm on the class hierarchy. To avoid cycles, visited classes are marked. Once a suitable templates is selected all classes along the search path are annotated with their distance to the selected candidate and a reference to the selected template. These annotations can be used to prune the search space of subsequent iterations. To optimize the overall process for resources with the same sets of classes we maintain a dictionary of the hashed sorted list of class IRIs and the selected template throughout the rendering process. Using this dictionary no resource that is subsequently rendered with the same set of classes needs to initiate a new template selection process.

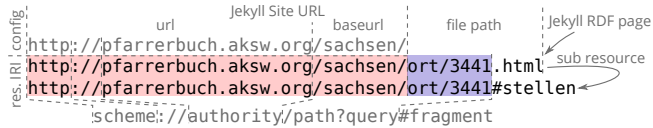

**Figure 3.** Scheme of the selection of page paths based on the configured Jekyll Site URL.

### 3.2 Resource Mapping and Site Creation

Jekyll RDF is meant to provide an HTML view for Linked Data resources and thus we need to map the resource's HTTP IRIs to file system paths, which is depicted in fig. 3. Taking an RDF resource Jekyll RDF matches it with the Jekyll Site URL as it is set in lines 1 and 2 of the configuration in listing 1. The matched part is than removed from the resource IRI and the remaining part `ort/3441` is mapped to the file path `ort/3441.html`. A query string in an IRI can be treated in the same way as a path. A fragment part is interpreted and truncated by the browser during a request, thus the second resource IRI ends up at the same IRI as the previous one. As fragment identifiers are usually used to identify anchors in HTML pages, all IRIs with a fragment are added to a list data structure (`page.subResources`) of the respective non fragment IRI page. All resources that do not match the configured base IRI are rendered with a, currently unstable, fallback solution under the directory `rdfsites`.

It is possible that a resource created by Jekyll RDF would overwrite a page that was already generated by Jekyll or any other plugin. In this case Jekyll RDF merges the two page object into one page and copies the content of the previously existing page into the content variable (cf. section 3.3) of the template selected for the RDF resource. If a template is specified for the Jekyll page it overwrites the template selected for the RDF resource.

### 3.3 Jekyll RDF Markup Language

The Jekyll RDF Markup Language (JML) is based on Liquid and extends it by filters to access the RDF graph from within a template. Liquid provides the concepts of *objects*, *tags*, and *filters*[20]. *Objects* define placeholders in a template

**Listing 2.** A simple template with HTML markup enriched by JML filters.

```
1   <h1>{{ page.rdf | rdf_property: "rdfs:label", "en" }}</h1>
2   <div>{{ page.rdf | rdf_property: "dct:created" | date: "%Y-%m-%d" }}</div>
3
4   {% assign publicationlist = "ex:publicationlist" | rdf_container %}
5   <ul>
6   {% for pub in publicationlist %}
7   <li>{{ pub | rdf_property: "dc:title" }}</li>
8   <li>{{ pub | rdf_property: "dct:creator", false, true | join: ", " }}</li>
9   {% endfor %}
10  </ul>
```

**Table 1.** The filters defined by Jekyll RDF and the tasks performed by them.

| Filter | Parameters ([optional]) | Description |
|---|---|---|
| rdf_get | | Get a variable representing a resource from the RDF graph. |
| rdf_property | IRI, [language, bool] | Get a value of a property of a resource (lines 1, 2, and 7 of listing 2). If the last parameter is set to true an array is returned (line 8). |
| rdf_inverse_property | IRI, [bool] | Get the value of an inverse property. |
| rdf_collection and rdf_container | [IRI] | Get RDF collections and RDF containers from the RDF graph as shown in line 4 of listing 2. |
| sparql_query | sparql query | Execute a SPARQL Query on the RDF graph, the passed value is bound to ?resourceUri or to ?resourceUri_n if an array is provided. |

to insert values and are denoted by two curly braces `{{ … }}`. The special object `{{content}}` in templates is a placeholder for the content of the rendered page. *Tags* are used to embed control flow into the template, they are denoted by curly braces and percent signs `{% … %}`. The tag `{% assign = "some value" %}` is used to assign a value to a variable. *Filters* manipulate the output of objects or the value of variables, they are chained and applied from left to right and separated by a vertical bar `|`. A filter gets a *value* passed from the left and can get *parameters* passed to the right after a colon `:` and separated by commas.

On every page which is generated by Jekyll RDF the variable `page.rdf` is present to reference the RDF resource represented by that page. To provide a way to access the RDF graph from within the template Jekyll RDF defines new Liquid filters as shown in table 1 and listing 2. The usage of Liquid filters allows to chain existing filters and filters defined by further plugins to the output of the JML filters. The JML filters accept a resource IRI as string or a resource object to be passed as value, they are shown and described in table 1.

## 4 Integration With Data Publication Workflows

With our approach we follow the single responsibility principle: *Make each program do one thing well* [12]. This principle is recently gaining attention with the increase of the importance of *micro services* to manage complex software systems. Following this principle it is possible to integrate Jekyll RDF with existing tools, workflows, management systems, and interfaces to build a full SCMS or to support data engineers to publish RDF graphs as HTML pages. A pragmatic and in software engineering already proven successful approach for coordinating collaboration and exchange of artifacts is the usage of Git repositories. The Quit Store [2,3] is a method on top of Git to version and collaboratively manage RDF knowledge repositories. In the following we show two aspects to consider when integrating Jekyll RDF with data management and content management workflows. We present two setups that adapt the continuous integration method from software engineering to build flexible data publication workflows with Jekyll RDF in section 4.1. To close the gap between structured data and human accessible browsing interfaces based on Jekyll RDF it is equally important to make the underlying RDF data available. We discuss possibilities to integrate the HTML and RDF publication with each other in section 4.2.

### 4.1 Using Jekyll RDF with a Continuous Integration

Continuous Integration (CI) is a concept used in software engineering to automatically test, build, and deploy software artifacts. This concept recently increases in usage for data engineering [19]. With Jekyll RDF it is possible to define a step in a CI system to render and deploy pages whenever the data is updated. Travis CI[21] is a hosted continues integration service used to build and test software projects at GitHub. Using a continuous integration system during the work in a team allows to produce automated builds and feedback during the development process whenever a contributor updates the templates or data in the Git repository. In fig. 4 we show two possible setups of automatic continuous deployment pipelines to publish Jekyll RDF sites. The setup in fig. 4a shows a completely publicly hosted setup that uses the Travis CI service to build the Jekyll RDF site and the webspace provided by GitHub pages to serve the produced Static Site. This setup allows a flexible collaboration system combined with a free of charge deployment without the need to maintain a complex infrastructure. The setup in fig. 4b is slightly complexer and allows the differentiation between a stable "master" version and an unstable "develop" version. In combination with Docker it is possible to build ready to deploy system images including the rendered site. Whenever an updated system image is available the deployment restarts the respective service with the updated image.

### 4.2 Integration with RDF Publication

Following the Linked Data principle: *When someone looks up a URI, provide useful information, using the standards (RDF\*, SPARQL)*[22] one page is created for each resource in the knowledge graph. Each resource page *provides useful information* in a human accessible format. Since the web page is build based on RDF data, besides the HTML representation, it is also important to make the underlying RDF data available to Semantic Web agents. To achieve this, various methods exist to make the respective data available as needed. In the following we discuss possibilities that have an overlap with the setup of Static Sites.

---

[21] `https://travis-ci.org`
[22] `http://www.w3.org/DesignIssues/LinkedData.html`

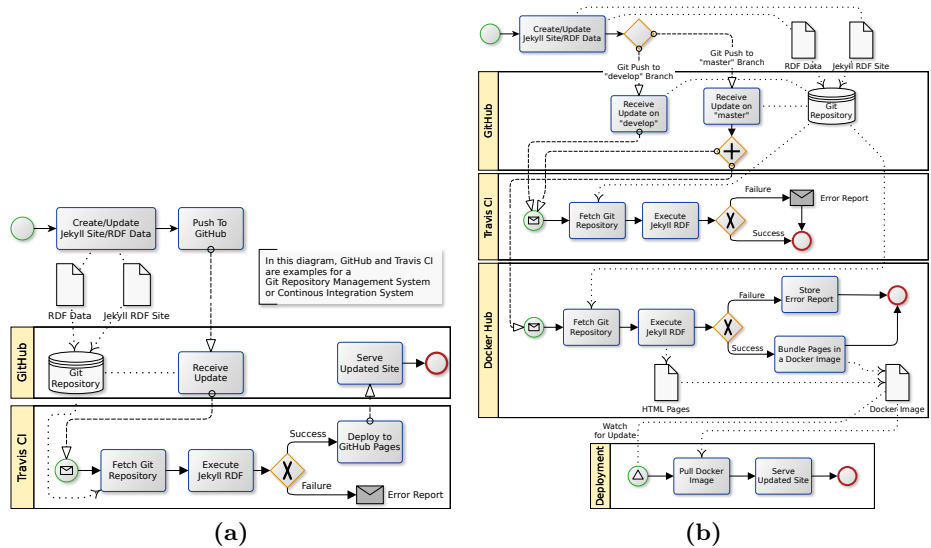

**Figure 4.** Jekyll RDF integrated in two different deployment pipelines.

A way to embed data in a document is to use RDFa properties within the HTML tags. Since RDFa interferes with the surrounding HTML-tags it is not subject to Jekyll RDF. Instead the embedding has to be performed by the template designer. To support the designer this problem could also be subject to future work for building a *Jekyll RDFa* plugin on top of Jekyll RDF. Instead of embedding the RDF data each resource can also be served in a separate file. In this case a script is employed to process the same data source as Jekyll RDF and produce individual Linked Data resource pages. To provide a possibility for an agent to find the respective resource page a link can be set or the HTTP server employs content negotiation. But in many cases, where Static Sites are used, the user has no ability to interfere with the HTTP protocol of the server. In this case the only way is to add an *HTML header link tag*[23] to point to the resource page, this can be embedded directly in any Jekyll template. If the HTTP protocol of the server can be configured the same link can be set as *HTTP Link header*[24]. In this case also *content negotiation*[25] is an elegant commonly used way to serve the correct representation of the resource based on the request.

---

[23] https://www.w3.org/TR/2017/REC-html52-20171214/document-metadata.html#the-link-element
[24] http://www.rfc-editor.org/rfc/rfc5988.txt
[25] https://tools.ietf.org/html/rfc7231#section-5.3, https://www.w3.org/DesignIssues/Conneg

## 5  Application

In the following we show a selection of use cases where we have successfully applied Jekyll RDF to build customized data exploration interfaces. The use cases are from different domains and provide very individual user interfaces as shown in figs. 5 and 6. The first two use cases are from the Digital Humanities, followed by a Current Research Information System, and finally we eat our own dogfood and present the usability of Jekyll RDF to build vocabulary description pages. In table 2 we compare the setups according to the number of triples, defined templates and pages, resulting pages, and execution time.

**Digital Humanities** The *Multilingual Morpheme Ontology* (MMoOn; [15]) is an ontology in the linguistics to create language resources of morphemic data (inventory) for inflectional languages. With the Open Hebrew Inventory we created a language specific extension of the MMoOn vocabulary for the Modern Hebrew language. The dataset currently consists of 197, 374 RDF Statements describing Hebrew words, lexemes, and morphs. Using the Jekyll RDF system we could create an interface to the Open Hebrew dataset that is specifically adapted to the presentation of Hebrew language data. To define the templates we created a Git repository[26] that consists of four class templates, two overview pages, and the dataset. On each update to the Git repository the Travis CI (cf. section 4.1) executes the Jekyll RDF process to create the 13, 404 inventory pages, 20 vocabulary pages, and two overview pages. Figure 5a shows the exploration interface at the example of the word הֻכְתַּב (hukhtav). Next to the HTML interface we created RDF resource pages in Turtle and RDF/JSON serialization. This RDF representation of the data is used to attach a system for users to contribute to the data through Structured Feedback [1]. In this way further dynamic elements can be provided in a static page based on the RDF data.

Another Project from the Digital Humanities is the *Pfarrerbuch* project[27]. In this project we build a Research Data Management system for historians to create a database of all pastors who served in Saxony, Hungary, the Church Province of Saxony, and Thuringia since the reformation. Especially in the filed of history we see a great need for customized and easily accessible user interfaces.

**Smart Data Analytics Work Group** The Smart Data Analytics research group (SDA[28]) investigates machine learning techniques (*analytics*) using structured knowledge (*smart data*). Machine learning requires sufficient data as training datasets. SDA investigated techniques which could help also to build such a dataset to depict the organizational structure and entities representing SDA. The SDA knowledge graph contains entities about persons, groups, projects, and publications as well as their relations. It is used for question answering, faceted

---

[26] http://mmoon-project.github.io/JekyllPage/
[27] https://github.com/AKSW/pfarrerbuch.jekyllrdf
[28] http://sda.tech, https://github.com/SmartDataAnalytics/sda.tech

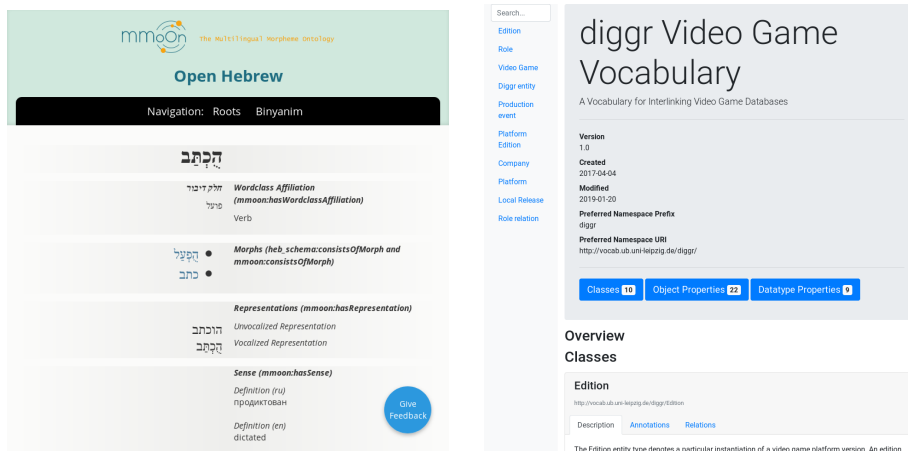

**(a)** MMoOn exploration interface showing the example of the resource for the Hebrew word הֻכְתַּב (hukhtav).

**(b)** The documentation page for the diggr OWL vocabulary used to model the interplay of games and publishers in global game culture research.

**Figure 5.** Images of Jekyll RDF pages to depict the variety of usage scenarios.

**Table 2.** Comparison of the presented Jekyll RDF setups. The average runtime is measured at the Travis CI over the last 10 runs.

| Setup | #Triples | #Templates/#Pages | #Res. Pages | Avg. Runtime |
|---|---|---|---|---|
| Open Hebrew Inventory | 197,374 | 4/2 | 13,426 | 1952.4 s (32.54 min) |
| Pfarrerbuch (demo subset) | 1,685 | 4/- | 138 | 8 s |
| SDA Work Group | 27,295 | 8 (Class) + 49 (Inst.)/2 | 253 | 327.3 s (5.45 min) |
| diggr Vocabulary | 221 | 1/- | 1 | 8 s |

search, data mining, and analysis for better decision making while hiring or restructuring the group. Using Jekyll RDF and the Linked Data principles helps to reuse the existing knowledge graph to build the work group homepage. In this way it is possible to publish a Current Research Information System (CRIS) with Jekyll RDF based on the SDA knowledge graph on the Web as shown in fig. 6.

**Vocabulary Documentation** One of the major advantages of using RDF to describe resources is the universality of the model and the ability to describe the schema resources as part of the same graph as the described data. This enables consumers and producers of the data to use an exploitable, executable, and metadata rich knowledge framework. eccenca[29] is a European enterprise based in Germany with a strong vision how semantic vocabularies and Linked Data can be used to integrate project data and data management. An important aspect of semantic data integration are vocabularies which capture essential concepts from the customers domain. These vocabularies build the foundation for a semantic data landscape which interlinks all customer datasets into an exploitable graph. In order to communicate the results of an ontology specification work, it is necessary to visualize and document the vocabularies. Existing ontology documentation tools lack the ability to extensively customize the result. Using Jekyll RDF we could build a set of templates for all major ontology classes and publish it as a theme called Jekyll Ontology Documentation project (JOD)[30]. However, fetching OWL constructs is problematic with SPARQL based graph access alone and also complex functionality such as the generation of a Manchester Syntax description is currently missing and should be integrated soon. It can easily be integrated in an existing RDF vocabulary project as shown at the example of the diggr Video Game Vocabulary project[31] performed by the Leipzig University Library. The deployment of Jekyll RDF with the JOD theme and a CI/CD pipeline in order to create the vocabulary documentation was straight forward. The user interface of the vocabulary documentation is shown in fig. 5b.

---

[29] http://www.eccenca.com

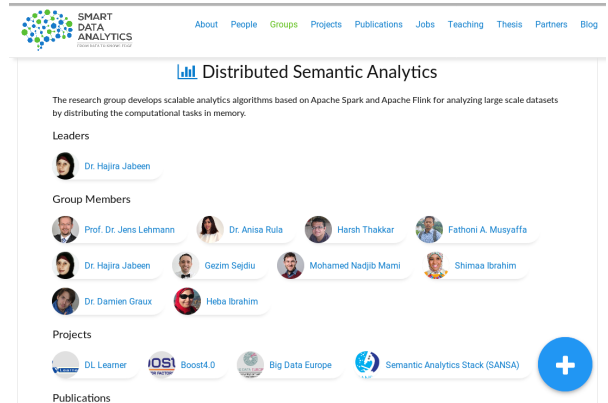

**Figure 6.** A work group page showing the members and projects.

## 6 Conclusion and Future Work

With the presented Jekyll RDF system we provide a methodology to close the gap between RDF data publishing and highly customizable publication and exploration interfaces. With our system it is possible to separate the management of RDF dataset from the creation of appropriate templates for the data presentation. Because we piggyback on the successful concept of Static Site Generators a low entry barrier is provided and the system requirements for the HTTP server are low. There is no need for costly dynamic server side computations which often also rely on the availability of a hard to maintain SPARQL endpoint. As shown in section 5 and table 2 the system allows the quick publication of small RDF dataset like RDF vocabulary, but also the creation of pages for huge datasets of more then 10*k* pages is possible with just a view templates. Especially, for the publication of highly interlinked datasets the usage of Jekyll RDF has assets as shown by the CRIS use case. As a Static Site Generator performs the work of creating the HTML pages in advance, the creation and the serving can be separated. The use of computing power is predictable and not affected by the amount of page visits. The separation allows a maximum flexibility in scaling the delivery of the site. It is possible to make extensive use of caching mechanisms such as *content delivery networks* to reduce the workload on the server and increase the availability of a site. In contrast to caching of dynamic pages the maintenance of static sites does not suffer from the problem of cache invalidation which lowers the effort of the publication.

---

[30] https://github.com/eccenca/jod, https://rubygems.org/gems/jekyll-theme-jod

[31] https://diggr.github.io/diggr-video-game-vocabulary/

In contrast to the related work of LD-R and MIRA [14,7] and Gatsby we provide a template based approach that aims at users without software developing experience. With the JML we minimize the effort of publishing Linked Data without the need to write a single line of programming code. Using JML as domain specific language allows also Web Designers to integrate knowledge from RDF graphs into their work. The template based approach is similar to our previous work with LESS and the OntoWiki Site Extension [5,10,9]. However, the previous work as well as the application frameworks presented in [14,7] relies on a complex dynamic data management back-end and SPARQL endpoint. With Jekyll RDF we present a static approach to maximize the scalability as it is independent of the availability of dynamic components at runtime.

As we extended Jekyll for our prototype we can benefit from the big ecosystem of plugins to enrich the created site. For the future work the performance of the generation process can be improved by an incremental build process to reuse pages from previous builds. To increase the usability of the presented method as a SCMS a set of predefined themes to be used with Jekyll RDF can support users, as shown by JOD. Looking at the possibilities of this concept in combination with the successful and generic design of RDF we see a great potential for future use cases. Due to the plethora of Static Site Generators we hope to see implementations to adopt our conception and methods to further systems like Next, Hugo, and Hyde[15]. There is no need to decide whether to publish data or a customized human readable interface anymore as the can be server next to each other on a static webspace.

## Acknowledgements

Thanks to the 2016 *Software Technik Praktikum* course group who did the initial implementation of Jekyll RDF: Elias Saalmann, Christian Frommert, Simon Jakobi, Arne Jonas Präger, Maxi Bornmann, Georg Hackel, Eric Füg. This work was partly supported by grants from the German Federal Ministries of Education and Research (BMBF) for the LEDS Project (03WKCG11C, 03WKCG11A),